\documentclass[11pt]{article}

\usepackage{acl}

\usepackage{times}
\usepackage{latexsym}
\usepackage[T1]{fontenc}

\usepackage[utf8]{inputenc}

\usepackage{microtype}

\usepackage{inconsolata}

\usepackage{graphicx}

\usepackage{booktabs}
\usepackage{multicol}
\usepackage{multirow}
\usepackage{kotex}
\usepackage{xcolor}
\usepackage{amsmath}
\usepackage{enumitem}
\usepackage{amssymb}

\newcommand{\ie}{{\it i.e.,}}
\newcommand{\eg}{{\it e.g.,}}
\newcommand{\ours}{BAGEL}

\title{Bayesian Active Learning with Gaussian Processes Guided by LLM Relevance Scoring for Dense Passage Retrieval}

\author{
    \textbf{Junyoung Kim}$^1$\thanks{Work done while visiting the University of Toronto.}, \textbf{Anton Korikov}$^2$, \textbf{Jiazhou Liang}$^2$, \textbf{Justin Cui}$^2$, \\
    \textbf{Yifan Simon Liu}$^2$, \textbf{Qianfeng Wen}$^2$, \textbf{Mark Zhao}$^2$ and \textbf{Scott Sanner}$^2$\thanks{Corresponding author.} \\[1ex]
    $^1$Sungkyunkwan University, Republic of Korea, $^2$University of Toronto, Canada \\
    \texttt{junyoung44@skku.edu}
}

\begin{document}
\maketitle

%
%
%
\begin{abstract}

While Large Language Models (LLMs) exhibit exceptional zero-shot relevance modeling, their high computational cost necessitates framing passage retrieval as a \textit{budget-constrained global optimization} problem. Existing approaches passively rely on first-stage dense retrievers, which leads to two limitations: (1) failing to retrieve relevant passages in semantically distinct clusters, and (2) failing to propagate relevance signals to the broader corpus.
To address these limitations, we propose \textit{\textbf{B}ayesian \textbf{A}ctive Learning with \textbf{G}aussian Processes guid\textbf{E}d by \textbf{L}LM relevance scoring (\textbf{\ours})}\footnote{ \url{https://github.com/junieberry/BAGEL}}, a novel framework that propagates sparse LLM relevance signals across the embedding space to guide global exploration. \ours\ models the multimodal relevance distribution across the entire embedding space with a query-specific Gaussian Process (GP) based on LLM relevance scores. Subsequently, it iteratively selects passages for scoring by strategically balancing the exploitation of high-confidence regions with the exploration of uncertain areas.
Extensive experiments across four benchmark datasets and two LLM backbones demonstrate that \ours\ effectively explores and captures complex relevance distributions and outperforms LLM reranking methods under the same LLM budget on all four datasets.
\end{abstract}

\section{Introduction}
\begin{figure*}[t]
\centering
\includegraphics[width=1\linewidth]{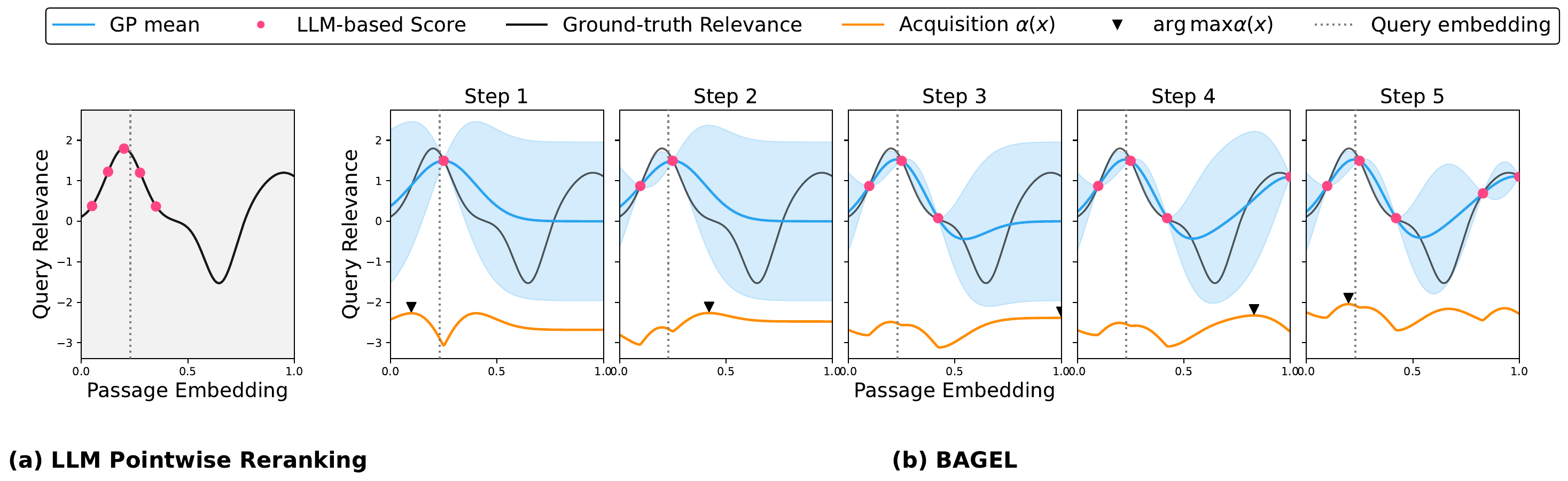}
\caption{Comparison of passage selection strategies under a fixed LLM budget. The x-axis shows a 1D projection of passage embeddings, and the y-axis displays the estimated Gaussian Process posterior mean (blue line) and standard deviation (shaded region). (a) LLM Pointwise Reranking focuses on passages (red dots) located nearby the query embedding (dashed line), often missing relevant passages in distant semantic clusters. (b) \ours~actively explores the embedding space with active learning. By modeling the predictive mean (blue line) and uncertainty (blue shaded area) from observed scores (red dots), the acquisition function (orange line) guides the selection of the next passage (black ▼) to explore high uncertainty regions and/or exploit high expected relevance areas, uncovering relevant passages in diverse clusters.}
\label{fig:bo}
\end{figure*}

While Large Language Models (LLMs) demonstrate a strong zero-shot ability in modeling complex query-passage relevance~\cite{LLM:qg, LLM:rankgpt}, their high computational cost prohibits exhaustive inference over large corpora. Consequently, the passage retrieval task can be framed as a \textit{budget-constrained global optimization problem}: identifying relevant passages within a massive search space under a strictly limited number of LLM inferences.
Prevalent approaches largely adhere to the \emph{LLM reranking} paradigm~\cite{LLM:yesno,LLM:rankgpt}, where dense retrievers retrieve a top-$K$ candidate set based on computationally efficient vector similarity and then reorder the set with an LLM.  
However, as shown in Figure~\ref{fig:bo}a, passively relying on this first-stage candidate set introduces two fundamental limitations:

\textbf{First}, the first-stage retriever imposes an upper bound on the recall. Relevant passages often reside in multiple, semantically distinct clusters scattered throughout the embedding space~\cite{cluster:diverse,liu2025ma}, forming a multimodal relevance distribution. These clusters may be located far from the query due to out-of-distribution domains~\cite{drlimit:unseen} or query ambiguity~\cite{drlimit:diversify}. Standard dense retrievers, which retrieve from a local neighborhood around the query embedding, may fail to detect global structures or high-relevance clusters located far from the initial neighborhood of query.

\textbf{Second}, existing approaches fail to propagate relevance signals from previously scored passages to unseen passages, overlooking the underlying embedding space that connects them. By failing to utilize the semantic structure where the relevance of one passage often implies the relevance of its neighbors~\cite{cluster:ir}, current approaches hinder efficient exploration of the global landscape.

To overcome these limitations, we propose Gaussian Process (GP)-based active learning as an effective framework for this budget-constrained task. GPs offer two intrinsic properties that directly address the aforementioned challenges:

\textbf{(1) \textit{Kernel-based Relevance Signal Propagation}}: GPs naturally model correlations between data points via kernel functions, enabling the representation of multimodal, clustered relevance functions (cf. Figure~\ref{fig:bo}b). 
They also allow us to interpolate relevance signals across the embedding space, effectively inferring the relevance of unobserved passages based on sparse LLM signals.

\textbf{(2) \textit{Uncertainty for Active Learning}}: GPs provide a probabilistic posterior that includes both predictive mean and variance (uncertainty). This enables Bayesian Active learning to effectively search the embedding space by balancing exploration (\eg\ via uncertainty) with exploitation.

Building on these insights, we propose \textit{\textbf{B}ayesian \textbf{A}ctive learning with \textbf{G}aussian Processes guid\textbf{E}d by \textbf{L}LM relevance scoring (\textbf{\ours})}, by leveraging the advantages of GPs to actively navigate the embedding space.  
Figure~\ref{fig:bo}b illustrates how \ours\ models relevance scores and actively selects the next passage for LLM scoring using both the GP predicted mean (blue line) and uncertainty (blue shaded area) as further elaborated below.  

Specifically, \ours\ constructs a query-specific GP surrogate model of the LLM-based query-passage relevance function, defined over the dense passage embedding space. By conditioning on the observed LLM relevance scores, \ours\ extends relevance signals to unseen passages, capturing complex, multimodal relevance structures and producing a full ranking under a limited LLM budget.  

Moreover, to 
explore multiple high relevance clusters (\ie\ modes), \ours\ actively selects additional passages (Steps 1--5 in Figure~\ref{fig:bo}b) for LLM relevance scoring using an acquisition function. By jointly evaluating the GP's predicted relevance and its uncertainty, it balances \textit{exploitation}: selecting passages predicted to be highly relevant, and \textit{exploration}: selecting passages with high uncertainty. This strategy enables efficient navigation of the \emph{entire} passage embedding space while making parsimonious use of expensive LLM relevance scoring.

The main contributions of our work are summarized as follows:
\begin{itemize}[leftmargin=1em, itemsep=0.2em, topsep=0.2em]
    \item \textbf{Gaussian Process-based Active Learning for Passage Exploration.} We propose a framework that integrates LLM-based relevance scoring with active learning driven by Gaussian Processes, enabling the strategic exploration of the dense passage embedding space.
    
    \item \textbf{Empirical Analysis of Kernel and Acquisition Functions.} We empirically show that stationary kernels (\eg\ RBF, Matérn) are effective for capturing multimodal relevance structures, while uncertainty-based acquisition functions play a critical role in guiding effective exploration.
        
    \item \textbf{Comprehensive Validation.} Evaluated on four distinct passage retrieval datasets, \ours\ significantly outperforms conventional LLM reranking baselines under the same LLM budget, for example, improving NDCG@50 from 29.3 to 41.6 on the TravelDest dataset.

\end{itemize}

\begin{figure*}[t]
\centering
\includegraphics[width=1\linewidth]{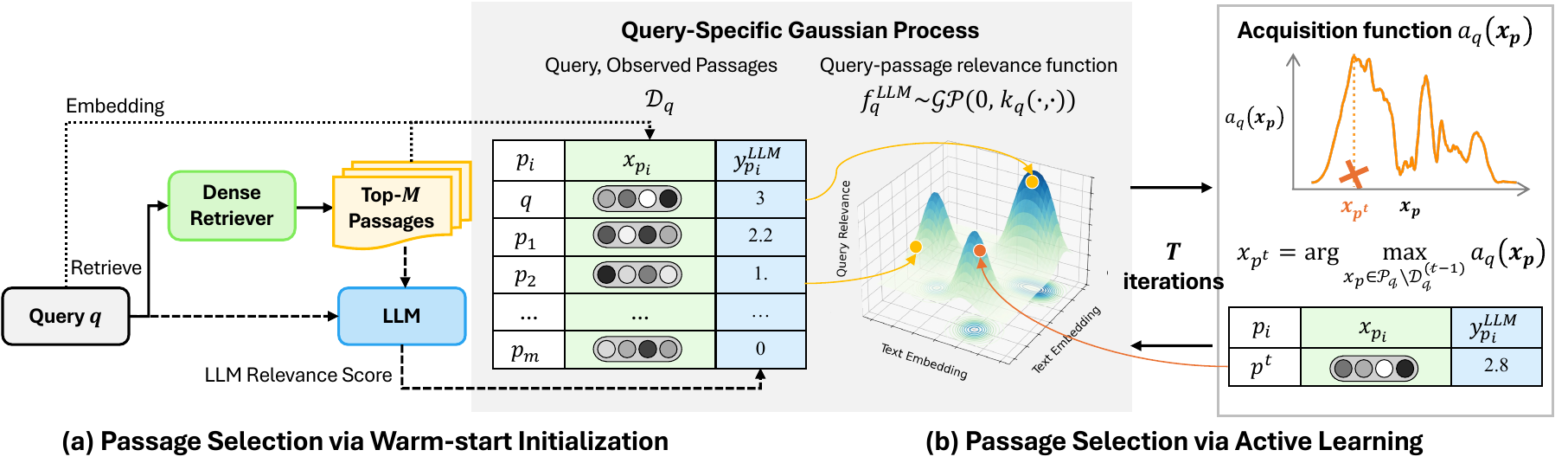}
\caption{Overview of \ours. For each query, \ours~defines a query-specific Gaussian Process (GP) using LLM-based relevance scores of selected passages in dense embedding space. The process begins with a (a) warm-start initialization by labeling the query itself and top-$M$ dense-retrieved passages. In the (b) active learning phase, an acquisition function combines the GP’s predictive mean and uncertainty to iteratively select next passage.}
\label{fig:overview}
\end{figure*}
\section{Preliminaries}

\subsection{Gaussian Processes}
\label{sec:gp}

GPs are non-parametric Bayesian models that define a prior over functions, allowing probabilistic function modeling\cite{gp:gpr, gp:gpml}. Let \( f: \mathbb{R}^d \rightarrow \mathbb{R} \), then the prior over \( f \) is specified as 
\begin{equation}
f \sim \mathcal{GP}(0, k(\cdot, \cdot)),
\end{equation} 
where \( k: \mathbb{R}^d \times \mathbb{R}^d \rightarrow \mathbb{R} \) is a chosen kernel function that aims to determine how similarity is measured and defines the function space the GP can represent. Consequently, the choice of kernel function influences properties such as smoothness and complexity of the GP. 
Given a set of inputs \( \mathbf{X} = [\mathbf{x}_1, \ldots, \mathbf{x}_n]^\top \in \mathbb{R}^{n \times d} \) with an observed output \( \mathbf{f}(\mathbf{X})=\mathbf{y} \in \mathbb{R}^n \), the posterior predictive distribution for a new input \( \mathbf{x}_* \in \mathbb{R}^d \) follows a Gaussian distribution with mean and uncertainty:
\begin{equation}
\label{eq:gp_mean}
\mu(\mathbf{x}_*) = \mathbf{k}_*^\top (\mathbf{K} + \alpha \mathbf{I})^{-1} \mathbf{y},
\end{equation} 
\begin{equation}
\label{eq:gp_var}
\sigma^2(\mathbf{x}_*) = k(\mathbf{x}_*, \mathbf{x}_*) - \mathbf{k}_*^\top (\mathbf{K} + \alpha \mathbf{I})^{-1} \mathbf{k}_*,
\end{equation} 
where \( \mathbf{K} \in \mathbb{R}^{n \times n} \) is the kernel matrix with \( \mathbf{K}_{ij} = k(\mathbf{x}_i, \mathbf{x}_j) \), \( \mathbf{k}_* \in \mathbb{R}^n \) contains the kernel values between \( \mathbf{x}_* \) and $\mathbf{x}$ (\ie\ $k( \mathbf{x}_*,\mathbf{x})$), and \( \alpha > 0 \) is a hyperparameter that accounts for observation noise. 
The value of \( f(\mathbf{x}_*) \) is estimated by the posterior mean \( \mu(\mathbf{x}_*) \), while \(\sigma^2(\mathbf{x}_*)\) quantifies the epistemic uncertainty of the model.

\subsection{Acquisition Function}
\label{sec:acq}

To guide the sequential search process effectively, it is essential to utilize both the predictive mean and the associated uncertainty provided by the GP. This is achieved through an acquisition function, which maps the posterior distribution to a utility value of each input:
\begin{equation}\label{eq:acq_general}
    a(\mathbf{x}) = \phi\big(\mu(\mathbf{x}),\, \sigma(\mathbf{x});\, \boldsymbol{\theta}\big).
\end{equation}
Here, \(\phi(\cdot)\) defines the specific strategy for combining the posterior mean and variance. By iteratively maximizing the acquisition function and updating the GP posterior, the process actively navigates the search space to converge toward the global optimum.

\subsection{LLM Query-Passage Relevance Scoring}
\label{sec:llm_based_score}
Recent work~\cite{LLM:yesno, LLM:qg, LLM:prp, LLM:zeroretriever} has proposed using LLMs to estimate query-passage relevance in a zero-shot setting. Following previous work~\cite{LLM:yesno}, we formulate the relevance estimation as a constrained generation task where the output is restricted to a single token representing a relevance score (\eg\ the token ``1''). Under this formulation, given a query \(q\), a passage \(p\), and an instruction prompt \texttt{prompt}, the LLM produces a vector of logits \( z = [z_0, z_1, \dots, z_{K-1}] \), where each component corresponds to a predefined integer relevance label \( r_k \in \{0, 1, \dots, K-1\} \).

\begin{equation} \mathbf{z} = \text{LLM}(\texttt{prompt}, q, p) \in \mathbb{R}^K. \end{equation}

To derive a scalar relevance score from these logits, \cite{LLM:yesno} introduces two variants for scoring. The first, \textit{expected relevance} (ER), interprets the logits as a probability distribution via the softmax function and computes the expected relevance value:

\begin{equation}\label{eq:er} S_{\text{ER}}(q, p) = \sum_{k=0}^{K-1} \underbrace{\left( \frac{e^{z_k}}{\sum_{j=0}^{K-1} e^{z_j}} \right)}_{P(r_k \mid q, p)} r_k. \end{equation}

The second variant, \textit{peak relevance} (PR), assigns the score corresponding to the relevance label with the highest logit:
\begin{equation}\label{eq:pr}
    S_{\text{PR}}(q, p) = r_{\arg\max_k z_k}.
\end{equation}
We will empirically examine the performance of the proposed method using both variants across datasets in Section~\ref{sec:exp_results}.
\section{\ours: Bayesian Active Learning with Gaussian Processes Guided by LLM Relevance Scoring}

In this section, we present \ours, a passage retrieval framework that integrates dense retrieval, LLM relevance scoring, and GP–guided exploration. \ours\ employs a query-specific GP to capture the query-passage relevance distribution based on a set of passages with observed LLM relevance scores. These passages are iteratively selected through an active selection process that balances exploration and exploitation as derived from the GP posterior estimates. We begin by introducing the query-specific GP, followed by two phases of active passage selection: (i) a warm-start initialization phase and (ii) an exploration phase guided by an acquisition function.

\subsection{Query-Specific Gaussian Process with LLM Relevance Score}
We first define the query-specific GP for each query \(q\), based on LLM relevance scores and extending the formulation in Section~\ref{sec:gp}. The query-specific GP takes as input the dense encoder embeddings of the passage \(p\), \(\mathbf{x}_p \in \mathbb{R}^d\), and the query \(q\), \(\mathbf{x}_q \in \mathbb{R}^d\), along with their observed LLM relevance scores, to model the query-passage relevance function \(f_q^{\mathrm{LLM}}(\mathbf{x}_p)\). This model can then be used to estimate the relevance score for any passage \(p_*\) with an unknown LLM relevance score.

Formally, for a query \(q\), let 
\[
\mathcal{D}_q = \{ (\mathbf{x}_{p_i},\, y^{\mathrm{LLM}}_{p_i}) \}_{i=1}^n
\]
denote the set of observed passages with known LLM relevance scores, where \(\mathbf{x}_{p_i} \in \mathbb{R}^d\) is the dense embedding of passage \(p_i\). 
Let \(S(\cdot, \cdot)\) be the LLM relevance scoring function defined in Section~\ref{sec:llm_based_score} (\eg\ \(S_{\text{ER}}\) or \(S_{\text{PR}}\)). 
For each passage \(p_i\), the score 
\[
y^{\mathrm{LLM}}_{p_i} = S(q, p_i)
\] 
represents its LLM relevance score with respect to the query. The query-specific GP for the query-passage relevance function is then defined as 
\begin{equation}\label{eq:gp}
    f_q^{\mathrm{LLM}} \sim \mathcal{GP}\big(0,\, k(\cdot, \cdot)\big),
\end{equation}
where \(k\) is the kernel function chosen as discussed in Section~\ref{sec:gp}.

While our framework supports any valid kernel function, we adopt the Radial Basis Function (RBF) kernel as the default choice in our experiments. The RBF kernel~\cite{gp:gpml} is a standard choice of GP and is defined as
\begin{equation}
\label{eq:rbf}
k_{\text{RBF}}(\mathbf{x}, \mathbf{x}') = \exp\left(-\frac{\|\mathbf{x} - \mathbf{x}'\|^2}{2\ell^2}\right),
\end{equation}
where the length-scale \(\ell\) determines how quickly correlations decay with distance. This kernel assumes that inputs that are closer in the input space produce more strongly correlated outputs, allowing the GP to represent complex functions. Additionally, we examine Linear and Matérn kernels, detailed in Appendix~\ref{sec:app_kernel}, with empirical comparisons provided in Section~\ref{sec:result_kernel}.

Building on the preliminaries, the posterior predictive distribution for a passage \(p_*\) with dense embedding \(\mathbf{x}_{p_*}\) follows a Gaussian distribution with mean and variance given by:  
\begin{equation}\label{eq:mu}
\mu_q(\mathbf{x}_{p_*}) = \mathbf{k}_*^\top \big(\mathbf{K} + \alpha \mathbf{I}\big)^{-1} \mathbf{y}^{\mathrm{LLM}},
\end{equation}  
\begin{equation}\label{eq:sigma}
\sigma_q^2(\mathbf{x}_{p_*}) = k(\mathbf{x}_{p_*}, \mathbf{x}_{p_*}) - \mathbf{k}_*^\top \big(\mathbf{K} + \alpha \mathbf{I}\big)^{-1} \mathbf{k}_*,
\end{equation}  
where \(\mathbf{K} \in \mathbb{R}^{n \times n}\) now represents the kernel matrix between the dense embeddings of the selected passages in $\mathcal{D}_q$, and \(\mathbf{k}_* \in \mathbb{R}^n\) is the kernel vector between the previously observed passages and the selected passage \(\mathbf{x}_{p_*}\).

\subsection{Passage Selection via Warm-start Initialization}

Our warm-start strategy aims to initialize \ours\ with reliable, high-quality signals in embedding regions that are likely to contain relevant passages. This approach reduces early-stage uncertainty and mitigates the cold-start problem. To this end, we utilize the query itself and top-ranked passages from a dense retriever to guide the initial selection process.

We first treat the query itself as an observation with maximum relevance. \ours\ treats the query embedding \(\mathbf{x}_q\) as an observation and assigns it a relevance score \(y_q^{\mathrm{LLM}}\) equal to the maximum possible value under the defined LLM relevance scoring function (\eg\ \(S_{\text{ER}}\) or \(S_{\text{PR}}\)). This ensures that the GP receives a strong positive signal and guided to model the relationship between the query and passages.
To augment this initial signal with promising passages, we retrieve the top-$M$ passages $\{p_1, p_2, \dots, p_M\}$ using a dense retriever given the query \(q\).
The resulting observed pairs $\{(\mathbf{x}_{p_i}, y^{\mathrm{LLM}}_{p_i})\}_{i=1}^M$, together with the query embedding–score pair $(\mathbf{x}_q, y_q^{\mathrm{LLM}})$, form the warm-start observation set:
\begin{equation}
        \mathcal{D}_q^{(0)} = \{ (\mathbf{x}_{p_i}, y_{p_i}^{\mathrm{LLM}}) \}_{i=1}^M \cup \{ (\mathbf{x}_q, y_q) \}.
\end{equation}

\subsection{Passage Selection via Active Learning}

After warm-start initialization, \ours\ iteratively selects new passages for LLM relevance scoring by applying the \emph{acquisition function} introduced in Section~\ref{sec:acq}. 
This phase enables efficient exploration of the entire passage embedding space while judiciously allocating expensive LLM relevance scoring.

Instead of generic inputs, we evaluate the utility of each passage embedding $\mathbf{x}_{p}$ using the query-specific GP predictive mean $\mu_q(\mathbf{x}_{p})$ and predictive standard deviation $\sigma_q(\mathbf{x}_{p})$.
While \ours\ is compatible with various strategies, we primarily utilize the Upper Confidence Bound (UCB)~\cite{gp:gpucb}.


Specifically, the UCB acquisition function is defined as
\begin{equation}\label{eq:acq_ucb}
    a^{\mathrm{UCB}}(\mathbf{x}) = \mu_q(\mathbf{x}) + \sqrt{\beta} \, \sigma_q(\mathbf{x}),
\end{equation}
where $\beta > 0$ is a hyperparameter that balances exploration (high uncertainty) and exploitation (high predictive mean).
Additional acquisition function formulations are provided in Appendix~\ref{sec:app_acq}, and empirical comparisons can be found in Section~\ref{sec:result_acq}.

At iteration $t \in \{1, \ldots, T\}$, where \(T\) denotes the remaining LLM budget, \ours\ selects the next passage \(p^t\) to label by maximizing the acquisition score over the pool of unlabeled passages:
\begin{equation}
    \mathbf{x}_{p^t} = \arg\max_{\mathbf{x}_{p} \in \mathcal{P} \setminus \mathcal{D}_q^{(t-1)}} a^{\mathrm{UCB}}(\mathbf{x}_{p}),
\end{equation}
where \(\mathcal{P}\) is the set of all passages and \(\mathcal{D}_q^{(t-1)}\) is the set of passages with observed relevance scores up to the previous iteration \(t-1\).
We then obtain the LLM-based relevance score $y^{\mathrm{LLM}}_{p^t} = S(q, p^t)$ and update the observation set:
\begin{equation}
    \mathcal{D}_q^{(t)} = \mathcal{D}_q^{(t-1)} \cup \big\{ (\mathbf{x}_{p^t},\, y^{\mathrm{LLM}}_{p^t}) \big\}.
\end{equation}

This process repeats for \(T\) iterations. Finally, the updated GP estimates the relevance scores for all passages in \(\mathcal{P}_q\). Notably, \ours\ supports any-time prediction; the query-specific GP can generate a ranking over all passages after any iteration $t$, making the method highly adaptable to online settings with varying budget constraints.

\begin{table*}[t]
\centering
\small
\caption{Overall Performance with a LLM budget of 50 per query. DR and CE denote the dense retriever and cross encoder baselines, respectively. List. and Point. indicate Listwise and Pointwise LLM. Bold denotes the best score within the same LLM backbone. R@k and N@k refer to Recall@k and NDCG@k, respectively.}
\setlength{\tabcolsep}{4.5pt}
\begin{tabular}{c|crrr|rrrrr|rrrrr}
\toprule
\multirow{3}{*}{Dataset} & LLM & \multicolumn{3}{c|}{Baseline} & \multicolumn{5}{c|}{Qwen3-14B} & \multicolumn{5}{c}{GPT-4o} \\ \cmidrule{2-15} 
 & \multirow{2}{*}{\shortstack{Score/ \\Method}} & \multicolumn{1}{c}{\multirow{2}{*}{BM25}} & \multicolumn{1}{c}{\multirow{2}{*}{DR}} & \multicolumn{1}{c|}{\multirow{2}{*}{CE}} & \multicolumn{1}{c|}{\multirow{2}{*}{List.}} & \multicolumn{2}{c|}{PR} & \multicolumn{2}{c|}{ER} & \multicolumn{1}{c|}{\multirow{2}{*}{List.}} & \multicolumn{2}{c|}{PR} & \multicolumn{2}{c}{ER} \\
 &  & \multicolumn{1}{c}{} & \multicolumn{1}{c}{} & \multicolumn{1}{c|}{} & \multicolumn{1}{c|}{} & \multicolumn{1}{c}{Point.} & \multicolumn{1}{c|}{Ours} & \multicolumn{1}{c}{Point.} & \multicolumn{1}{c|}{Ours} & \multicolumn{1}{c|}{} & \multicolumn{1}{c}{Point.} & \multicolumn{1}{c|}{Ours} & \multicolumn{1}{c}{Point.} & \multicolumn{1}{c}{Ours} \\ \midrule
\multirow{4}{*}{Covid} & N@10 & 50.4 & 57.0 & 66.5 & \multicolumn{1}{r|}{71.8} & 74.2 & \multicolumn{1}{r|}{76.6} & 76.5 & \textbf{77.2} & \multicolumn{1}{r|}{73.2} & 71.9 & \multicolumn{1}{r|}{70.4} & 74.6 & \textbf{74.7} \\
 & N@50 & 42.8 & 48.7 & 51.1 & \multicolumn{1}{r|}{52.3} & 52.9 & \multicolumn{1}{r|}{61.4} & 53.6 & \textbf{63.6} & \multicolumn{1}{r|}{52.6} & 52.4 & \multicolumn{1}{r|}{57.7} & 52.8 & \textbf{62.1} \\
 & R@10 & 1.7 & 2.0 & 2.3 & \multicolumn{1}{r|}{2.4} & 2.5 & \multicolumn{1}{r|}{\textbf{2.6}} & \textbf{2.6} & \textbf{2.6} & \multicolumn{1}{r|}{2.5} & 2.5 & \multicolumn{1}{r|}{2.5} & \textbf{2.6} & \textbf{2.6} \\
 & R@50 & 7.3 & 7.7 & 7.7 & \multicolumn{1}{r|}{7.7} & 7.7 & \multicolumn{1}{r|}{9.5} & 7.7 & \textbf{9.7} & \multicolumn{1}{r|}{7.7} & 7.7 & \multicolumn{1}{r|}{9.2} & 7.7 & \textbf{9.9} \\ \midrule
\multirow{4}{*}{NFCorpus} & N@10 & 31.5 & 31.2 & 33.7 & \multicolumn{1}{r|}{34.7} & 37.5 & \multicolumn{1}{r|}{37.7} & 37.5 & \textbf{38.4} & \multicolumn{1}{r|}{38.4} & 37.8 & \multicolumn{1}{r|}{38.9} & 39.1 & \textbf{40.6} \\
 & N@50 & 27.7 & 29.0 & 30.7 & \multicolumn{1}{r|}{30.9} & 32.7 & \multicolumn{1}{r|}{32.8} & 32.7 & \textbf{33.6} & \multicolumn{1}{r|}{33.0} & 32.8 & \multicolumn{1}{r|}{32.8} & 33.5 & \textbf{35.9} \\
 & R@10 & 15.1 & 15.3 & 15.5 & \multicolumn{1}{r|}{16.8} & 17.7 & \multicolumn{1}{r|}{18.0} & 17.5 & \textbf{18.3} & \multicolumn{1}{r|}{18.1} & 17.9 & \multicolumn{1}{r|}{18.2} & 18.2 & \textbf{19.0} \\
 & R@50 & 21.4 & 25.3 & 25.3 & \multicolumn{1}{r|}{25.3} & 25.3 & \multicolumn{1}{r|}{25.9} & 25.3 & \textbf{26.4} & \multicolumn{1}{r|}{25.3} & 25.3 & \multicolumn{1}{r|}{24.5} & 25.3 & \textbf{27.3} \\ \midrule
\multirow{4}{*}{Robust04} & N@10 & 38.9 & 39.1 & 46.4 & \multicolumn{1}{r|}{50.5} & 52.8 & \multicolumn{1}{r|}{55.9} & 53.8 & \textbf{57.3} & \multicolumn{1}{r|}{50.6} & 55.5 & \multicolumn{1}{r|}{60.4} & 57.3 & \textbf{62.1} \\
 & N@50 & 34.9 & 33.2 & 36.1 & \multicolumn{1}{r|}{36.8} & 38.2 & \multicolumn{1}{r|}{44.4} & 38.6 & \textbf{45.7} & \multicolumn{1}{r|}{35.2} & 39.1 & \multicolumn{1}{r|}{45.9} & 40.0 & \textbf{48.7} \\
 & R@10 & 13.4 & 12.1 & 14.2 & \multicolumn{1}{r|}{14.3} & 15.6 & \multicolumn{1}{r|}{16.7} & 15.7 & \textbf{17.4} & \multicolumn{1}{r|}{14.0} & 16.3 & \multicolumn{1}{r|}{17.6} & 16.4 & \textbf{18.8} \\
 & R@50 & 29.0 & 24.9 & 24.9 & \multicolumn{1}{r|}{24.9} & 24.9 & \multicolumn{1}{r|}{30.7} & 24.9 & \textbf{31.7} & \multicolumn{1}{r|}{21.7} & 24.9 & \multicolumn{1}{r|}{30.7} & 24.9 & \textbf{33.2} \\ \midrule
\multirow{4}{*}{TravelDest} & N@10 & 21.1 & 22.3 & 43.1 & \multicolumn{1}{r|}{48.6} & 45.8 & \multicolumn{1}{r|}{49.8} & 48.3 & \textbf{51.0} & \multicolumn{1}{r|}{53.3} & 48.9 & \multicolumn{1}{r|}{57.0} & 55.1 & \textbf{58.5} \\
 & N@50 & 17.2 & 21.6 & 26.4 & \multicolumn{1}{r|}{27.7} & 27.2 & \multicolumn{1}{r|}{37.4} & 27.9 & \textbf{38.0} & \multicolumn{1}{r|}{28.8} & 28.2 & \multicolumn{1}{r|}{40.2} & 29.3 & \textbf{41.6} \\
 & R@10 & 1.0 & 0.9 & 1.7 & \multicolumn{1}{r|}{2.1} & 2.1 & \multicolumn{1}{r|}{\textbf{2.5}} & 2.2 & \textbf{2.5} & \multicolumn{1}{r|}{2.3} & 2.0 & \multicolumn{1}{r|}{2.7} & 2.3 & \textbf{2.9} \\
 & R@50 & 3.4 & 3.9 & 3.9 & \multicolumn{1}{r|}{3.9} & 3.9 & \multicolumn{1}{r|}{6.7} & 3.9 & \textbf{6.8} & \multicolumn{1}{r|}{3.9} & 3.9 & \multicolumn{1}{r|}{6.7} & 3.9 & \textbf{6.9} \\ \bottomrule
\end{tabular}
\label{tab:main}
\end{table*}

\section{Experimental Setup}
We address the following research questions:

\begin{itemize}[leftmargin=3em, itemsep=0em, topsep=0em]
    \item[\textbf{RQ1}:] Does \ours\ outperform LLM reranking baselines under a fixed LLM budget across different LLMs?
    \item[\textbf{RQ2}:] How does \ours\ balance exploration and exploitation in the embedding space?
    \item[\textbf{RQ3}:] How do components such as the kernel and the acquisition function impact the performance of \ours?
\end{itemize}

\subsection{Baselines}
We compare \ours\ against five representative baselines spanning traditional sparse retrieval (\textbf{BM25}~\cite{ir:bm25}), dense retrieval (\textbf{Dense Retriever}~\cite{ir:sentencebert}), neural reranking (\textbf{Cross Encoder}~\cite{ir:multibert}), and LLM-based reranking (\textbf{Pointwise LLM}~\cite{LLM:yesno} and \textbf{Listwise LLM}~\cite{LLM:rankgpt}). We refer to Appendix~\ref{sec:app_baselines} for detailed descriptions of each baseline. We evaluate all methods using NDCG and Recall at cutoffs of $k=10, 50$.

\subsection{Datasets}
We evaluate our method on four passage retrieval datasets, including Covid, NFCorpus, and Robust04 from the BEIR benchmark~\cite{data:beir}, and TravelDest~\cite{data:traveldest1, data:traveldest2}. The former two datasets serve as domain-specific benchmarks, while the latter two feature ambiguous queries. Dataset statistics and details are reported in Appendix~\ref{sec:app_dataset}.

\subsection{Implementation Details}
\paragraph{\textbf{Budget}}
To ensure a consistent and fair comparison across ranking paradigms, we define the \textbf{budget} as the total number of individual passages evaluated by the LLM, rather than the raw count of API calls. Under this definition, Pointwise LLM and \ours\ each consume 1 unit per passage evaluation, as each call involves a single query-passage pair. For Listwise LLM, the budget is determined by the number of passages included in a single prompt; our RankGPT \cite{LLM:rankgpt} baseline employs a window size of 50, meaning one API call consumes 50 units.

For all methods, we fix a total budget of \textbf{50} per query, ensuring a fair comparison across methods regardless of how individual API calls are structured. For \ours, the budget is further partitioned into \(M=25\) for \textbf{warm-start initialization} and \(T=25\) for the \textbf{active learning} phase. Section~\ref{sec:budget} presents a comprehensive evaluation of the efficiency and effectiveness of \ours\ under these equivalent computational constraints.

\paragraph{\textbf{Settings}}
For \ours, we adopt an RBF kernel for the Gaussian Process, optimizing the length scale \(\ell\) with a learning rate of 0.01. We set the GP noise variance \(\alpha\) and UCB scaling factor \(\beta\) to 0.001 and 2, respectively, and use \texttt{all-MiniLM-L6-v2} (dimension 384) as the dense retriever backbone. For LLM backbones, we evaluate Qwen3-14B~\cite{imple:qwen} and GPT-4o~\cite{imple:gpt4o} across all methods. LLM query-passage relevance scoring for Pointwise LLM and \ours\ follows the Umbrela~\cite{imple:umbrela} prompting template, where each query-passage pair is assigned an integer label from \(\{0, 1, 2, 3\}\); the full prompt is provided in Table~\ref{tab:prompt}. Further details on \ours\ and baselines are provided in Appendix~\ref{sec:app_imp}.

\begin{figure*}[t]
\centering
\includegraphics[width=1\linewidth]{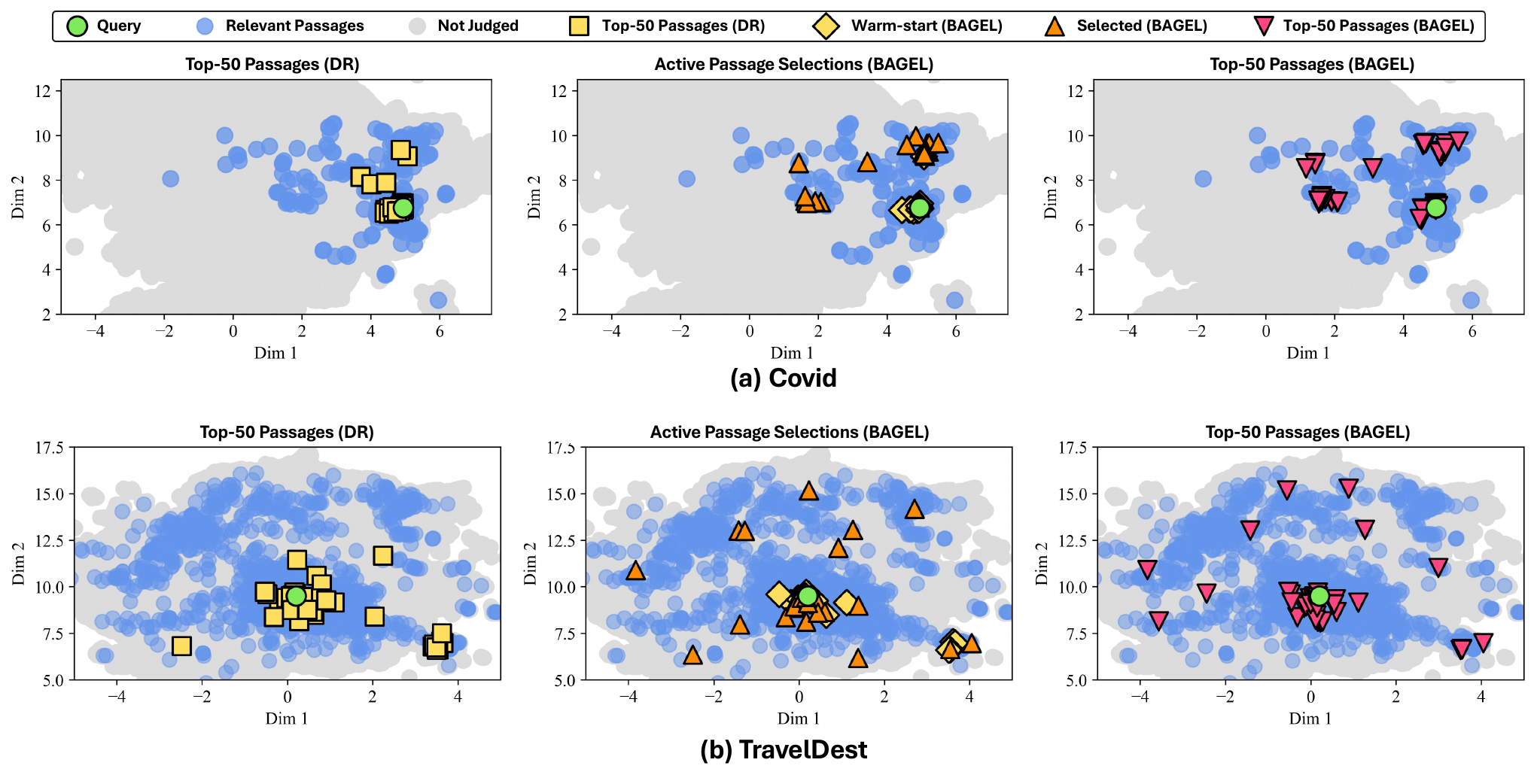}
\caption{UMAP projections from Covid and TravelDest showing how \ours~balances exploitation-exploration in two different relevance distributions: locally concentrated clusters ((a) Covid), and globally dispersed clusters ((b) TravelDest).  Observe specifically that \ours~(right) samples a broader selection of relevant passages (middle) compared to dense retrieval (left).}
\vspace{-2mm}
\label{fig:casestudy}
\end{figure*}


\section{Experimental Results}
\subsection{Overall Performance}\label{sec:exp_results}
To answer \textbf{RQ1}, Table~\ref{tab:main} presents the results on four datasets under a fixed LLM budget of 50 per query. Across all four datasets and both LLM backbones, \ours\ consistently outperforms all baselines methods.
Notably, in contrast to conventional rerankers confined to the initial candidate pool, the Recall@50 results demonstrate that \ours\ discovers relevant passages beyond this fixed set by actively exploring the embedding space.
Furthermore, the widening performance gap at deeper cutoffs (\eg\ NDCG@50 vs. NDCG@10) reveals that while baselines saturate on top-ranked positives, \ours\ utilizes these discovered passages to maintain high relevance density even at lower ranks where baselines typically falter.

\paragraph{\textbf{GPT-4o vs. Qwen3-14B}}
Table~\ref{tab:main} shows that GPT-4o not only boosts overall performance but also amplifies the gains of \ours\ over pointwise baselines compared to Qwen3-14B. We attribute this to GPT-4o's stronger ranking capability, which yields a relevance score distribution with lower noise. This high-quality supervision allows the Gaussian Process to learn a more stable posterior, thereby enhancing both mean estimation and uncertainty quantification for more effective exploration.

\paragraph{\textbf{ER vs. PR}}
Comparing the scoring functions, $S_{\mathrm{ER}}$ consistently yields larger gains over the pointwise LLM than $S_{\mathrm{PR}}$. This result stems from the inductive bias of GPs, which assumes a continuous, smooth target function. $S_{\mathrm{ER}}$ aligns with this assumption by providing smooth, real-valued scores via the probability-weighted average of relevance scores. In contrast, the coarse, discrete labels of $S_{\mathrm{PR}}$ result in step-like supervision, obscuring fine-grained relevance patterns and hindering effective posterior learning.

Based on these observations, subsequent experiments are conducted using $S_{\mathrm{ER}}$ with Qwen3-14B as the default backbone. The statistical significance of our results is validated through paired significance tests and bootstrap confidence intervals (see Appendix~\ref{sec:app_stats}). Further analysis on latency is provided in Appendix~\ref{sec:app_latency}.

\begin{figure}[t]
\centering
\includegraphics[width=1\linewidth]{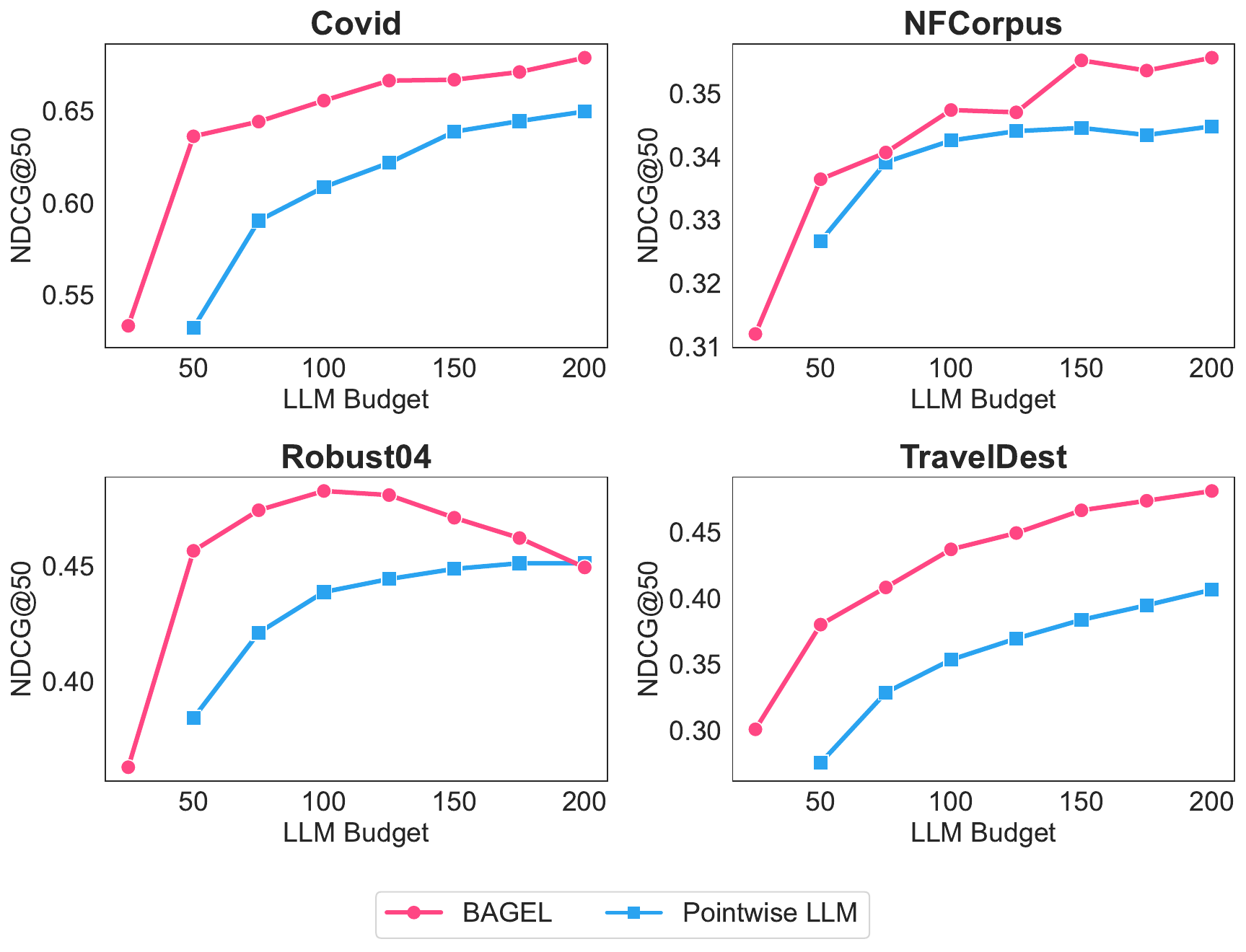}
\caption{Performance by different values of LLM budget.}
\label{fig:budget}
\end{figure}

\subsection{Budget}
\label{sec:budget}
To further examine whether \ours\ utilizes the LLM budget more effectively than existing LLM reranking methods, we compare performance under varying LLM budgets (\textbf{RQ1}), as illustrated in Figure~\ref{fig:budget}. The number of warm-start passages is fixed at 25, with the remainder allocated to active learning. As shown, \ours\ consistently outperforms the Pointwise LLM baseline. Notably, \ours\ achieves comparable performance with significantly fewer budget, demonstrating that it maximizes budget utility by propagating relevance signals via the GP model and strategically balancing exploitation and exploration.

\subsection{Case Study}
\label{sec:result_casestudy}
To better understand \ours's selection strategy for \textbf{RQ2}, Figure~\ref{fig:casestudy} shows UMAP~\cite{imple:umap} projections of (i) the dense retriever’s Top-50 passages, (ii) warm-start passages and passages selected during \ours's active learning, and (iii) Top-50 passages of \ours.

In Covid (Figure~\ref{fig:casestudy}a), for the query \textit{“what are the benefits and risks of re-opening schools in the midst of the COVID-19 pandemic?”}, the most relevant passages are concentrated near the query embedding, with a few located farther away. While the dense retriever overlooks a left-side cluster, \ours\ gradually explores high-uncertainty regions and uncovers these overlooked passages.

In TravelDest (Figure~\ref{fig:casestudy}b), for the general query \textit{“I want to capture stunning sunshine”}, the relevant passages are scattered across the entire embedding space. The dense retriever, constrained by its implicit unimodal assumption, focuses near the query embedding and neglects other clusters. In contrast, \ours\ actively explores diverse regions with high uncertainty and retrieves relevant passages from multiple clusters, leading to more balanced coverage.

These visualizations illustrate how \ours\ adapts its exploitation–exploration balance to the underlying relevance structure. In relatively concentrated distributions like Covid, it tends to explore locally, whereas in dispersed multimodal distributions such as TravelDest, it explores globally to cover all major clusters.

\begin{table}[t]
\footnotesize
\caption{Performance by different kernels with a LLM budget of 50 per query, where stationary kernels excel by capturing score multimodality.}
\centering
\setlength{\tabcolsep}{5pt}
\begin{tabular}{c|c|rrrr}
\toprule
\multicolumn{1}{c|}{Dataset} & Kernel & N@10 & N@50 & R@10 & R@50 \\ \midrule
\multirow{3}{*}{Covid} & Linear & 16.43 & 11.88 & 0.50 & 1.76 \\
 & Matérn & 75.24 & 62.31 & 2.54 & 9.64 \\
 & RBF & 77.24 & 63.64 & 2.60 & 9.66 \\ \midrule
\multirow{3}{*}{NFCorpus} & Linear & 36.91 & 31.95 & 17.35 & 24.74 \\
 & Matérn & 38.29 & 33.78 & 18.19 & 26.65 \\
 & RBF & 38.40 & 33.65 & 18.31 & 26.42 \\ \midrule
\multirow{3}{*}{Robust04} & Linear & 25.47 & 18.52 & 7.29 & 11.72 \\
 & Matérn & 57.71 & 46.23 & 17.32 & 32.26 \\
 & RBF & 57.28 & 45.66 & 17.39 & 31.72 \\ \midrule
\multirow{3}{*}{TravelDest} & Linear & 21.91 & 14.00 & 0.97 & 2.24 \\
 & Matérn & 50.92 & 38.04 & 2.53 & 6.72 \\
 & RBF & 50.96 & 38.02 & 2.52 & 6.82 \\ \bottomrule
\end{tabular}
\label{tab:kernel}
\end{table}

\subsection{Component Analysis}
We next examine how components influence the performance of \ours\ across datasets, addressing \textbf{RQ3}. Additional analyses on hyperparameters (\eg\ \(\alpha\), \(\beta\)) and the number of warm-start passages are provided in Appendix~\ref{sec:app_parameter}.

\paragraph{\textbf{Kernel}}
\label{sec:result_kernel}
To examine the impact of kernel choice, we compare the Linear and Matérn kernels in addition to the RBF kernel. Table~\ref{tab:kernel} shows results across four datasets. We observe that RBF and Matérn perform similarly and consistently outperform the Linear kernel. Unlike Linear, \emph{stationary kernels} (RBF and Matérn) effectively model complex, multimodal relevance landscapes by preserving local neighborhoods based on relative distance~\cite{gp:gpml}. Consequently, RBF and Matérn achieve comparable performance and significantly surpass the Linear kernel. This performance gap between two kernel types indicates the multimodality in the relevance score distribution. 


\begin{table}[t]
\footnotesize
\caption{Performance by different acquisition functions with a LLM budget of 50 per query, demonstrating the advantage of uncertainty-guided exploration over naive selection.}
\centering
\setlength{\tabcolsep}{4pt}
\begin{tabular}{c|c|rrrr}
\toprule
Dataset & Acq. & N@10 & N@50 & R@10 & R@50 \\ \midrule
\multirow{6}{*}{Covid} & Random & 76.0 & 60.4 & 2.5 & 9.1 \\
 & Dense & 73.9 & 57.9 & 2.4 & 8.4 \\ \cmidrule{2-6} 
 & PI & 76.9 & 60.7 & 2.6 & 9.3 \\
 & EI & 76.3 & 63.3 & 2.7 & 9.9 \\
 & TS & 75.8 & 62.0 & 2.5 & 9.4 \\
 & UCB & 77.2 & 63.6 & 2.6 & 9.7 \\ \midrule
\multirow{6}{*}{NFCorpus} & Random & 37.5 & 33.8 & 18.1 & 28.3 \\
 & Dense & 37.8 & 33.1 & 18.2 & 26.5 \\ \cmidrule{2-6} 
 & PI & 38.1 & 33.6 & 17.9 & 26.8 \\
 & EI & 38.5 & 33.8 & 18.0 & 26.9 \\
 & TS & 37.7 & 34.4 & 18.4 & 29.0 \\
 & UCB & 38.4 & 33.6 & 18.3 & 26.4 \\ \midrule
\multirow{6}{*}{Robust04} & Random & 50.2 & 40.8 & 14.8 & 28.7 \\
 & Dense & 54.3 & 41.6 & 15.7 & 28.5 \\ \cmidrule{2-6} 
 & PI & 57.1 & 45.8 & 16.9 & 32.0 \\
 & EI & 57.1 & 45.9 & 17.0 & 32.2 \\
 & TS & 50.4 & 41.5 & 14.6 & 29.2 \\
 & UCB & 57.3 & 45.7 & 17.4 & 31.7 \\ \midrule
\multirow{6}{*}{TravelDest} & Random & 44.7 & 34.1 & 2.1 & 5.8 \\
 & Dense & 49.9 & 36.3 & 2.4 & 6.0 \\ \cmidrule{2-6} 
 & PI & 51.9 & 38.4 & 2.6 & 6.8 \\
 & EI & 50.8 & 37.6 & 2.5 & 6.6 \\
 & TS & 42.4 & 33.7 & 2.0 & 5.8 \\
 & UCB & 51.0 & 38.0 & 2.5 & 6.8 \\ \bottomrule
\end{tabular}
\label{tab:acq}
\end{table}
\paragraph{\textbf{Acquisition Function}}
\label{sec:result_acq}
To evaluate active passage selection strategies, Table~\ref{tab:acq} compares several acquisition functions with 25 warm-start passages and 25 active selection passages. As a result Random and Dense yield the weakest performance, as they fail to balance exploration and exploitation: Random selects passages in an arbitrary manner, whereas Dense restricts itself to local optima. In contrast, Bayesian acquisition strategies (PI, EI, TS, and UCB) consistently outperform these baselines by leveraging the GP posterior to guide exploration. While differences among them are minor, they collectively demonstrate the advantage of uncertainty-guided selection.

\section{Related Work}

\subsection{Reranking Paradigm}
Passage retrieval aims to identify relevant passages from a large corpus given a query \cite{ir:introir}. Traditional approaches typically follow a multi-stage pipeline, starting with sparse lexical methods \cite{ir:bm25} or dense embedding models \cite{ir:dpr}. While scalable, these first-stage retrievers often struggle to capture complex semantic relationships. To address this, reranking paradigms \cite{ir:colbert, ir:multibert} employ more sophisticated models to reorder the candidate set retrieved in the initial stage.

Recently, Large Language Models (LLMs) have been integrated into this pipeline to further enhance relevance scoring through pointwise or listwise prompting. Pointwise methods score query-passage pairs independently using generation likelihood or fine-grained labels \cite{LLM:qg, LLM:yesno}, while listwise methods rank multiple passages simultaneously to capture inter-passage context \cite{LLM:zslistwise, LLM:rankgpt}. To mitigate the inefficiencies of these listwise methods, \citet{uncertaintyir:acurank} proposed adaptively allocating computation based on uncertainty estimation.

However, this pipeline is bottlenecked by the recall of initial retriever; any relevant passage missed in the first stage is lost for subsequent reranking. In contrast, \ours\ overcomes this constraint by utilizing uncertainty as a navigational guide to directly explore the entire embedding space, rather than being confined to a pre-filtered candidate list.

\subsection{Gaussian Processes}
Gaussian Processes (GPs) are non-parametric Bayesian models that define a distribution over functions, providing principled uncertainty quantification~\cite{gp:gpml, gp:gpr}. By modeling surrogate functions with well-calibrated uncertainty, GPs facilitate sample-efficient exploration via acquisition strategies~\cite{gp:gpucb}. This makes them particularly effective in limited-supervision scenarios, such as hyperparameter optimization~\cite{gp:hyper}, active learning~\cite{gp:activelearning}, and recommendation system tasks~\cite {liu2025multimodal}. Recent findings also demonstrate that GPs maintain robustness in high-dimensional settings~\cite{gp:highdim}. 

Despite their success in other domains, the application of Bayesian frameworks in retrieval has been largely limited to Bandit-style approaches~\cite{banditir:dynamicrag, banditir:mba}, which are primarily used to select among several predefined retrieval methods. Unlike these approaches, \ours\ departs from this paradigm by being the first to leverage GPs to treat the entire embedding space as the object of exploration. By modeling the query-specific relevance distribution across the continuous vector space, we utilize GP-based uncertainty as a navigational signal to extrapolate sparse LLM scoring results and enable global exploration under strict computational budgets.
\section{Conclusion}
In this paper, we introduce \ours, a framework that provides a novel integration of Gaussian Process-based Bayesian active learning with LLM relevance scoring. Unlike traditional reranking methods restricted to a static candidate set, \ours~actively explores the dense embedding space, propagating relevance signals to discover semantically distinct clusters that are often overlooked. Our experiments demonstrate that by leveraging Bayesian uncertainty to guide the selection of passages, \ours\ significantly improves retrieval performance on all four datasets compared to LLM reranking baselines under fixed computational budgets. Overall, \ours\ demonstrates the potential of Bayesian active learning combined with LLM-based relevance scoring to make effective and parsimonious use of a limited LLM budget for retrieval.


\section*{Limitations}
The limitations of this study can be primarily categorized into three aspects regarding data dependency, robustness, and scalability. First, the framework exhibits a strong dependence on the initial embedding quality, as the kernel function operates directly within the pre-trained dense retriever’s latent space; consequently, suboptimal semantic mapping or kernel mismatches may hinder the effective propagation of relevance signals across the manifold. Second, the sensitivity to LLM scoring noise poses a potential risk of reinforcing hallucinations. If the LLM generates factually incorrect relevance scores, the Gaussian Process may confidentially propagate these errors, leading to a degraded retrieval model that prioritizes misinformation. Finally, scalability to web-scale corpora remains a challenge, as calculating uncertainty for every unobserved document in a billion-scale index is computationally prohibitive, necessitating future research into efficient pruning or hierarchical search strategies for real-world deployment.

\section*{Acknowledgments}
This work was supported by the Institute of Information \& Communications Technology Planning \& Evaluation (IITP) grants funded by the Korea government (MSIT) (RS-2022-00143911, AI Excellence Global Innovative Leader Education Program; RS-2024-00457882, National AI Research Lab Project).

\bibliography{custom}

\clearpage
\appendix
\renewcommand{\arraystretch}{1.2}
\begin{table}[!h]
\small
\begin{tabular}{p{0.9\linewidth}}
\toprule
\textbf{Prompt} \\ \midrule
Given a query and a list of passages, you must provide a score on an integer scale of 0 to 3 with the following meanings:\\
0 = represent that the passage has nothing to do with the query,\\
1 = represents that the passage seems related to the query but does not answer it,\\
2 = represents that the passage has some answer for the query, but the answer may be a bit unclear, or hidden amongst extraneous information and\\
3 = represents that the passage is dedicated to the query and contains the exact answer.\\
Important Instruction: Assign category 1 if the passage is somewhat related to the topic but not completely, category 2 if passage presents something very important related to the entire topic but also has some extra information and category 3 if the passage only and entirely refers to the topic. If none of the above satisfies give it category 0.\\
Query: \{\textbf{query}\}\\
Passage: \{\textbf{passage}\}\\
Split this problem into steps:\\
Consider the underlying intent of the search.\\
Measure how well the content matches a likely intent of the query (M).\\
Measure how trustworthy the passage is (T).\\
Consider the aspects above and the relative importance of each, and decide on a final score (O). Final score must be an integer value only.\\
Do not provide any code or reasoning in result. Provide only the score without any explanation.\\
\#\#final score: \\
\bottomrule
\end{tabular}
\caption{Prompt for LLM-based scoring in Section~\ref{sec:llm_based_score}. Both \textbf{query} and \textbf{passage} are placeholders.}
\label{tab:prompt}
\end{table}

\section{Kernel Function}
\label{sec:app_kernel}
The kernel function \(k(\mathbf{x}, \mathbf{x}')\) defines the similarity between two inputs. We use the following kernels \cite{gp:kernel}:

\begin{itemize} [leftmargin=1.5em] 


    \item \textbf{Linear (Dot Product) Kernel:}
    \[
    k_{\text{lin}}(\mathbf{x}, \mathbf{x}')
    = \mathbf{x}^\top \mathbf{x}'
    \]

    \item \textbf{Matérn Kernel:}
    \[
    \begin{aligned}
    k_{\text{Matérn}}(\mathbf{x}, \mathbf{x}')
    &= \frac{2^{1-\nu}}{\Gamma(\nu)}
       \left(
         \frac{\sqrt{2\nu}\,\|\mathbf{x}-\mathbf{x}'\|}{\ell}
       \right)^{\nu} \\
    &\quad\times
       K_{\nu}\!\left(
         \frac{\sqrt{2\nu}\,\|\mathbf{x}-\mathbf{x}'\|}{\ell}
       \right)
    \end{aligned}
    \]

    \item \textbf{Radial Basis Function (RBF) Kernel:}
    \[
    k_{\text{RBF}}(\mathbf{x}, \mathbf{x}')
    = \exp\!\left(-\frac{\|\mathbf{x}-\mathbf{x}'\|^2}{2\ell^2}\right)
    \]

\end{itemize}

\section{Acquisition Functions}
\label{sec:app_acq}
Acquisition functions determine which passages to evaluate next. We employ two heuristic functions (Random, Dense) and four Bayesian functions (EI, PI, TS, UCB). Let $\mu(\mathbf{x})$ and $\sigma(\mathbf{x})$ denote the GP posterior mean and standard deviation for passage $\mathbf{x}$,  $f^*$ the current best observed score, and $\xi \ge 0$ a parameter controlling exploration. 
$\Phi(\cdot)$ and $\phi(\cdot)$ as the standard normal CDF and PDF.
\begin{itemize}

    \item \textbf{Random:}
    Selects passages uniformly at random without using any scoring function.
    
    \item \textbf{Dense:}
    Selects passages with the highest dense retriever scores, equivalent to selecting
    the top-$n$ passages from the initial warm-start set without active learning.
    
    \item \textbf{Probability of Improvement (PI)}~\cite{gp:pi}:
    Selects passages most likely to surpass $f^*$:
    \[
    \alpha_{\text{PI}}(\mathbf{x})
    = \Phi\!\left(
        \frac{\mu(\mathbf{x}) - f^* - \xi}{\sigma(\mathbf{x})}
      \right).
    \]

    \item \textbf{Expected Improvement (EI)}~\cite{gp:gpei}:
    Selects passages expected to yield the greatest improvement over $f^*$:
    \[
    \begin{aligned}
    \alpha_{\text{EI}}(\mathbf{x})
    &= (\mu(\mathbf{x}) - f^* - \xi)\,\Phi(Z)
       + \sigma(\mathbf{x})\,\phi(Z), \\
    Z &= \frac{\mu(\mathbf{x}) - f^* - \xi}{\sigma(\mathbf{x})}.
    \end{aligned}
    \]

    \item \textbf{Thompson Sampling (TS)}~\cite{gp:gpthompson}:
    Draws a sample from the GP posterior and selects the passage with the highest sampled value:
    \[
    \mathbf{x}_{\text{next}}
    = \arg\max_{\mathbf{x}} \tilde{f}(\mathbf{x}),
    \qquad
    \tilde{f} \sim \mathcal{GP}(\mu, k).
    \]

    \item \textbf{Upper Confidence Bound (UCB)}~\cite{gp:gpucb}:
    Selects passages with the highest optimistic estimate of relevance:
    \[
    \alpha_{\text{UCB}}(\mathbf{x})
    = \mu(\mathbf{x})
      + \sqrt{\beta}\,\sigma(\mathbf{x}),
    \]
    where $\beta > 0$ controls the exploration--exploitation trade-off.
\end{itemize}

\begin{table}
\footnotesize
\caption{Data statistics.}
\centering
\begin{tabular}{c|rrr}
\toprule
Dataset & \multicolumn{1}{l}{\# Query} & \multicolumn{1}{l}{\# Corpus} & \multicolumn{1}{l}{\# Qrel} \\ \midrule
Covid & 50 & 171,332 & 55,853 \\
NFCorpus & 323 & 3,633 & 12,288 \\
Robust04 & 249 & 528,155 & 308,857 \\
TravelDest & 100 & 131,268 & 2,314,420 \\ \bottomrule
\end{tabular}
\label{tab:dataset}
\end{table}

\section{Additional Experimental Setup}

\subsection{Baseline}
\label{sec:app_baselines}
We compare \ours\ against the following five baselines:

\begin{itemize}[leftmargin=*]
    \item \textbf{BM25}~\cite{ir:bm25}: A traditional sparse retrieval method based on term frequency\allowbreak–inverse document frequency (TF–IDF) weighting.
    
    \item \textbf{Dense Retriever}~\cite{ir:sentencebert}: A dual-encoder model that encodes queries and passages into dense vectors and retrieves based on vector similarity. This model serves both as the first-stage retriever for all reranking baselines and as the dense retrieval component for initializing \ours.
    
    \item \textbf{Cross Encoder}~\cite{ir:multibert}: A BERT-based reranker that jointly encodes a query–passage pair, leveraging token-level interactions to predict fine-grained relevance scores.
    
    \item \textbf{Pointwise LLM}~\cite{LLM:yesno}: An LLM-based method that independently scores each query–passage pair. Our implementation follows the scoring variants described in Section~\ref{sec:llm_based_score}. Note that \ours\ adopts this same approach to obtain LLM relevance scores during the active search process.
    
    \item \textbf{Listwise LLM}~\cite{LLM:rankgpt}: An LLM-based reranking approach that inputs a list of candidate passages into the LLM simultaneously to generate a reordered list based on global context.
\end{itemize}

\subsection{Dataset Statistics}
\label{sec:app_dataset}
Table~\ref{tab:dataset} reports the statistics of the datasets used in our experiments. TravelDest contains queries, passages, and cities, with each passage linked to a city in a many-to-one relationship. As it only provides query–city relevance annotations, we generate query–passage labels by prompting an LLM to assess the relevance between each query and passages from its ground-truth city. For this, we use Gemini-2.5-Flash~\cite{data:gemini} with a modified binary version of the Umbrela prompt~\cite{imple:umbrela}. Also, given the high number of relevant cities per query (average 113.58) of TravelDest, we labeled a wide range of passages (17.63\% of the candidate pool), resulting in a 3.11\% relevance rate. This extensive coverage ensures that the benchmark is both comprehensive and challenging. For each dataset, we list the number of queries, the size of the passage corpus, and the number of relevance annotations (qrels). These statistics are provided for completeness and reproducibility.

\begin{table}[t]
\footnotesize
\caption{Results of Paired Wilcoxon Signed-Rank Test. Qwen denotes Qwen3-14B. COV, NFC, ROB, and TRAV stand for COVID, NFCorpus, Robust04, and TravelDest, respectively. Statistically significant results ($p < 0.05$) are indicated in bold.}
\centering

\resizebox{0.97\linewidth}{!}{

\begin{tabular}{c|c|rrrr}
\toprule
LLM & Dataset & \multicolumn{1}{l}{N@10} & \multicolumn{1}{l}{N@50} & \multicolumn{1}{l}{R@10} & \multicolumn{1}{l}{R@50} \\ \midrule
\multirow{4}{*}{Qwen} & COV & 0.5720 & \textbf{0.0000} & 0.5310 & \textbf{0.0000} \\
 & NFC & \textbf{0.0472} & \textbf{0.0270} & \textbf{0.0035} & \textbf{0.0141} \\
 & ROB & \textbf{0.0000} & \textbf{0.0000} & \textbf{0.0000} & \textbf{0.0000} \\
 & TRAV & 0.1300 & \textbf{0.0000} & \textbf{0.0108} & \textbf{0.0000} \\ \midrule
\multirow{4}{*}{GPT-4o} & COV & 0.1100 & \textbf{0.0069} & \textbf{0.0390} & \textbf{0.0020} \\
 & NFC & \textbf{0.0006} & \textbf{0.0000} & \textbf{0.0001} & \textbf{0.0000} \\
 & ROB & \textbf{0.0000} & \textbf{0.0000} & \textbf{0.0000} & \textbf{0.0000} \\
 & TRAV & \textbf{0.0029} & \textbf{0.0000} & \textbf{0.0001} & \textbf{0.0000} \\ \bottomrule
\end{tabular}

}

\label{tab:paired_test}
\end{table}
\begin{table*}[t]
\footnotesize
\caption{Results of Bootstrap Confidence Intervals. Values in bold indicate intervals where both bounds are greater than zero.}
\centering
\begin{tabular}{c|c|rrrr}
\toprule
LLM & Dataset & \multicolumn{1}{c}{NDCG@10} & \multicolumn{1}{c}{NDCG@50} & \multicolumn{1}{c}{Recall@10} & \multicolumn{1}{c}{Recall@50} \\ \midrule
\multirow{4}{*}{Qwen3-14B} & Covid & [$-0.029$, $0.042$] & \boldmath{[$0.064, 0.139$]} & [$-0.001$, $0.001$] & \boldmath{[$0.011, 0.028$]} \\
 & NFCorpus & [$-0.006$, $0.024$] & [$-0.006$, $0.025$] & [$-0.005$, $0.021$] & [$-0.009$, $0.031$] \\
 & Robust04 & \boldmath{[$0.042, 0.093$]} & \boldmath{[$0.069, 0.108$]} & \boldmath{[$0.022, 0.041$]} & \boldmath{[$0.053, 0.085$]} \\
 & TravelDest & [$-0.007$, $0.060$] & \boldmath{[$0.079, 0.124$]} & \boldmath{[$0.001, 0.007$]} & \boldmath{[$0.019, 0.039$]} \\ \midrule
\multirow{4}{*}{GPT-4o} & Covid & [$-0.100$, $-0.006$] & \boldmath{[$0.014, 0.092$]} & [$-0.003$, $0.000$] & \boldmath{[$0.007, 0.027$]} \\
 & NFCorpus & \boldmath{[$0.004, 0.028$]} & \boldmath{[$0.012, 0.035$]} & [$-0.005$, $0.017$] & \boldmath{[$0.008, 0.042$]} \\
 & Robust04 & \boldmath{[$0.029, 0.070$]} & \boldmath{[$0.075, 0.108$]} & \boldmath{[$0.015, 0.033$]} & \boldmath{[$0.071, 0.100$]} \\
 & TravelDest & \boldmath{[$0.022, 0.090$]} & \boldmath{[$0.111, 0.154$]} & \boldmath{[$0.004, 0.012$]} & \boldmath{[$0.024, 0.045$]} \\ \bottomrule
\end{tabular}
\label{tab:bootstrap}
\end{table*}

\subsection{Additional Implementation Details}
\label{sec:app_imp}
All experiments were conducted using a single training run on an NVIDIA H100 GPU with a 40GB MIG partition.
\paragraph{\textbf{\ours}}
For \ours, The target values $y$ are standardized to have zero mean and unit variance prior to training.
Before each acquisition function evaluation, the Gaussian Process hyperparameters are reoptimized from scratch. 
The kernel length scale $\ell$ is optimized using a learning rate of $0.01$ and constrained to the interval $[0.01, 2]$, 
with a uniform prior defined over the same range, while the output scale is optimized using a $\mathrm{Gamma}(2, 2)$ prior. Note that \ours\ is deterministic under the UCB acquisition function, where the same set of observations guarantees an identical selection process and final ranking.

\paragraph{\textbf{Baselines}}
We implemented the retrieval baselines using standard open-source libraries: BM25 via Pyserini~\cite{code:pyserini} and dense retrievers/cross-encoders via \texttt{sentence-transformers}~\cite{ir:sentencebert}. To optimize computational efficiency during LLM inference, Qwen3-14B was loaded in 4-bit precision using the unsloth framework~\cite{imple:unsloth}. Regarding the LLM reranking baselines, pointwise LLM reranking is implemented identically to the query-passage relevance scoring of \ours. Full prompt templates used for pointwise scoring are detailed in Table~\ref{tab:prompt}. Listwise LLM ranking follows the RankGPT~\cite{LLM:rankgpt} with a sliding window size of 50.

\section{Statistical Analysis}
\label{sec:app_stats}

In this section, we provide a rigorous statistical evaluation to demonstrate that the performance gains achieved by \ours\ are both significant and robust. We employ two distinct methodologies: the Paired Wilcoxon Signed-Rank Test to verify the significance of improvements over baselines, and Bootstrap Confidence Interval (CI) analysis to assess the stability of these gains across various data distributions.

\subsection{Paired Statistical Significance}
First, we conducted the Paired Wilcoxon Signed-Rank Test to compare \ours\ against the strongest baselines. As shown in Table~\ref{tab:paired_test}, the majority of the tests, particularly for the GPT-4o experiments, yield $p$-values well below the $0.05$ threshold, confirming the statistical significance of \ours's superiority. Interestingly, the $p$-values for GPT-4o are consistently smaller and more stable across all datasets compared to Qwen3-14B. This aligns with our observation that GPT-4o's higher ranking consistency allows \ours\ to yield systematic improvements with minimal variance.

\subsection{Bootstrap Confidence Intervals}
To assess the stability of the improvements, we performed a bootstrap CI analysis. The results in Table~\ref{tab:bootstrap} show that while short-rank metrics (NDCG@10) vary across datasets, the 95\% CI for Recall remains significantly and stably positive in most cases. This reinforces our claim that \ours's active exploration of the embedding space effectively identifies diverse relevant clusters, leading to a more robust retrieval of the total relevant set.

\begin{table}
\footnotesize
\caption{Comparison of average latency (seconds per query) between \ours with different acquisition functions. Thomp. denotes Thompson Sampling and COV, NFC, ROB, and TRAV stand for COVID, NFCorpus, Robust04, and TravelDest.}
\centering
\begin{tabular}{ll|rrrr}
\toprule
\multicolumn{2}{l|}{Latency (sec)} & \multicolumn{1}{l}{COV} & \multicolumn{1}{l}{NFC} & \multicolumn{1}{l}{ROB} & \multicolumn{1}{l}{TRAV} \\ \midrule
\multicolumn{2}{l|}{Pointwise LLM} & 4.77 & 4.95 & 6.72 & 4.19 \\ \midrule
\multicolumn{1}{l|}{\multirow{6}{*}{Ours}} & Random & 5.48 & 5.10 & 8.62 & 4.94 \\
\multicolumn{1}{l|}{} & Greedy & 5.33 & 5.07 & 7.15 & 4.32 \\
\multicolumn{1}{l|}{} & PI & 7.44 & 6.17 & 12.98 & 6.83 \\
\multicolumn{1}{l|}{} & EI & 7.49 & 6.18 & 13.08 & 6.86 \\
\multicolumn{1}{l|}{} & Thomp. & 7.37 & 6.20 & 12.97 & 6.88 \\
\multicolumn{1}{l|}{} & UCB & 7.38 & 6.21 & 12.88 & 6.75 \\ \bottomrule
\end{tabular}
\label{tab:latency}
\end{table}

\section{Latency}
\label{sec:app_latency}
To assess computational efficiency, Table~\ref{tab:latency} compares the average latency per query of \ours\ against the LLM pointwise baseline. Although \ours\ requires additional time for GP updates, resulting in higher latency compared to the pointwise baseline, this overhead is acceptable given the substantial performance gains. Furthermore, both methods operate within the same constraint of 50 LLM calls, ensuring fair resource usage.

\section{Additional Parameter Study}
\label{sec:app_parameter}
\begin{figure}[t]
\centering
\includegraphics[width=1\linewidth]{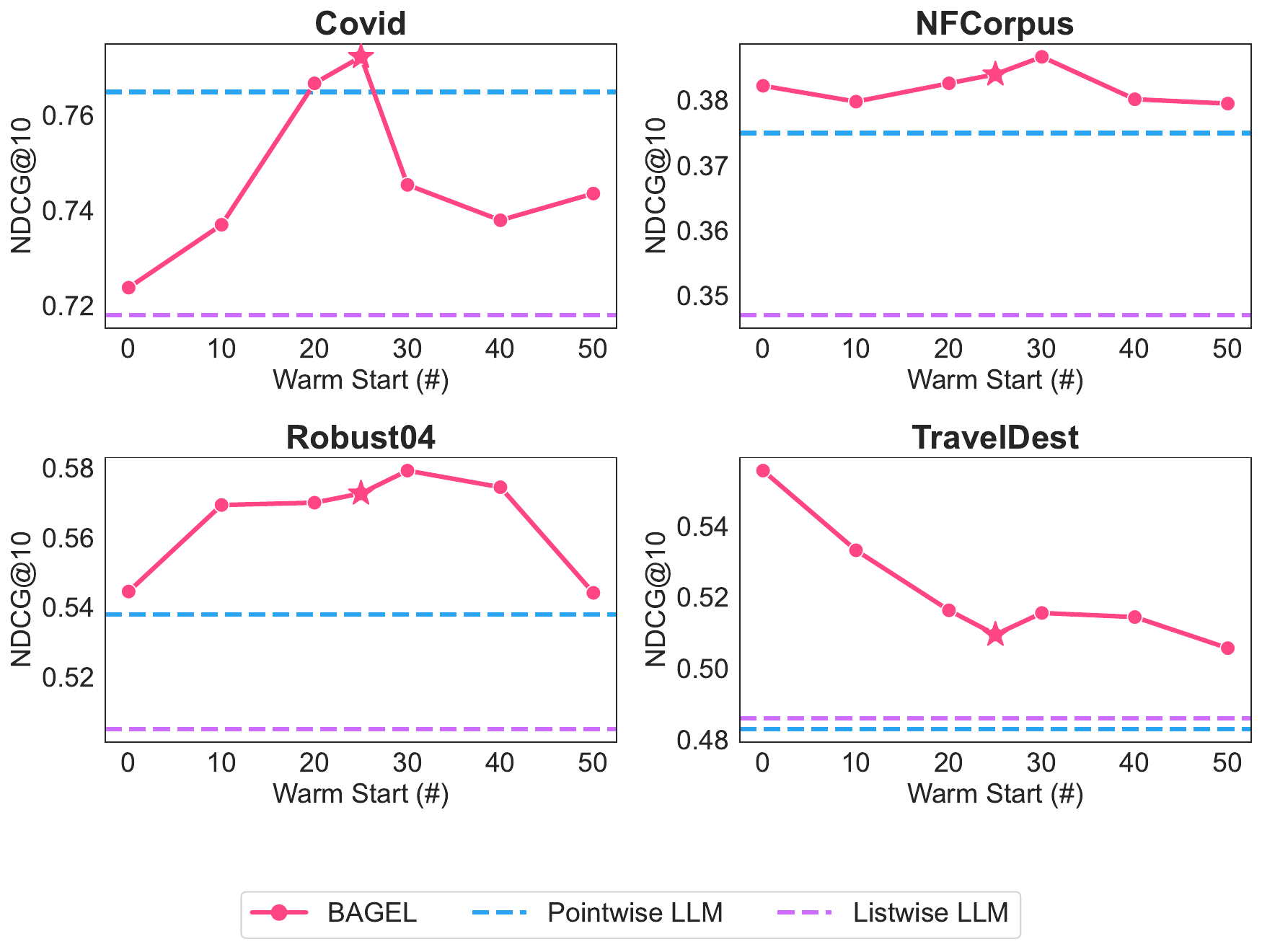}
\caption{Effect of number of warm-start passages on performance. The star represents performance with default value (\(M=25\)).}
\label{fig:warmstart}
\end{figure}

\begin{figure}[t]
\centering
\includegraphics[width=1\linewidth]{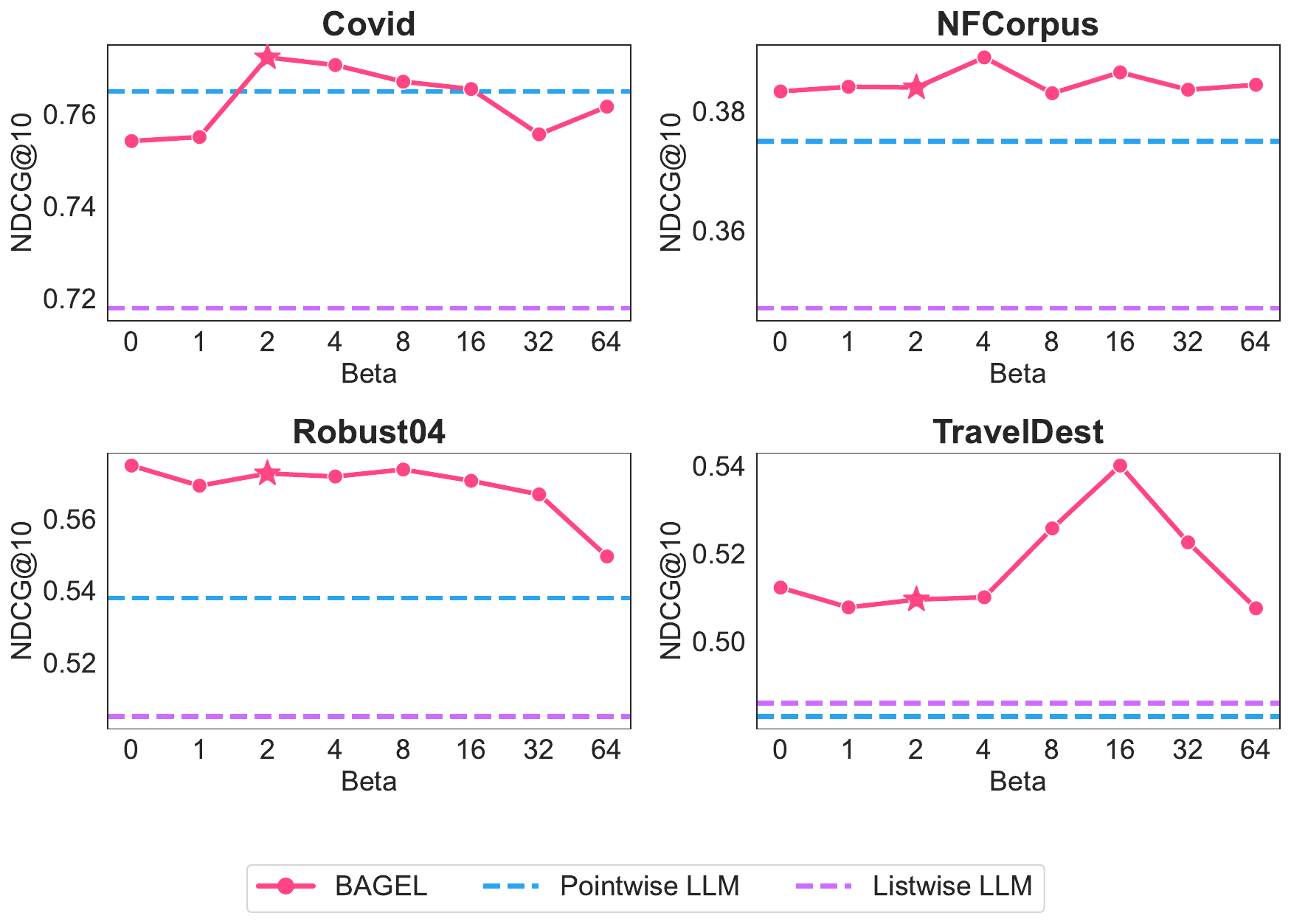}
\caption{Performance by different values of beta. The blue dots represent performance with default parameter (\(\beta=2\))}
\label{fig:beta}
\end{figure}

\begin{figure}[t]
\centering
\includegraphics[width=1\linewidth]{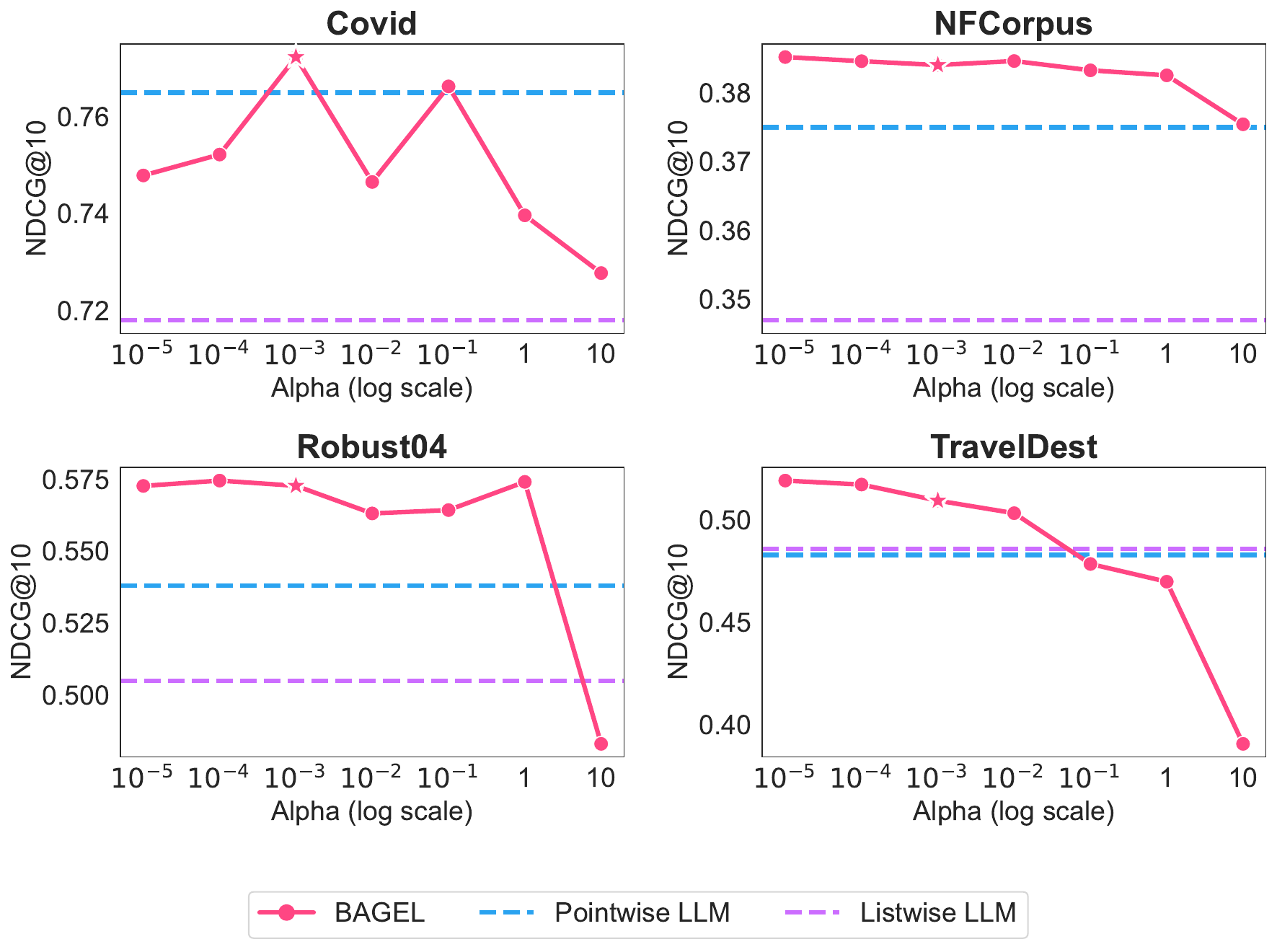}
\caption{Performance by different values of alpha. The star represent performance with default parameter (\(\alpha=1e-3\))}
\label{fig:alpha}
\end{figure}

\paragraph{\textbf{Number of Warm-start Passages}}
Figure~\ref{fig:warmstart} shows how varying the number of warm-start passages affects performance, with a fixed total LLM budget of 50 and the remainder after the warm start allocated to active learning. We observe different trends across datasets. This discrepancy may be due to differences in the quality of the warm-start passages retrieved by the dense retriever. In general, however, performance tends to be low when there is no warm start at all, likely because navigating the high-dimensional embedding space without any initial guidance is inherently challenging.

\paragraph{\textbf{Beta}}
Figure~\ref{fig:beta} examines the effect of the parameter $\beta$, which controls the balance between exploitation and exploration in UCB. A larger $\beta$ favors selecting samples with higher uncertainty, while a smaller $\beta$ prioritizes those with higher predicted relevance. The optimal value of $\beta$ varies across datasets, highlighting the importance of dataset-specific tuning. In cases where $\beta=0$ yields the best performance, the GP’s posterior mean may already provide sufficiently accurate predictions, making pure exploitation effective.

\paragraph{\textbf{Alpha}}
Figure~\ref{fig:alpha} shows that the optimal observation–noise parameter \(\alpha\) differs by dataset, reflecting variations in LLM score noise. While larger \(\alpha\) can help in noisier datasets, overly large values over-smooth the relevance function and hurt performance, making per-dataset tuning essential.

\end{document}